\documentclass{aa}

\usepackage[varg]{txfonts}
\usepackage{graphicx}

\usepackage{hyperref}
\usepackage{epstopdf}

\bibpunct{(}{)}{;}{a}{}{,} % to follow the A&A style

\usepackage{mathtools}

\usepackage{siunitx}
\usepackage[version=3]{mhchem}

\newcommand{\Msun}{\ensuremath{\mathrm{M}_{\odot}}}
\newcommand{\about}{\ensuremath{\mathord{\sim}}}	%no space behind tilde with \about
\newcommand{\ssup}[1]{\ensuremath{^{\mathrm{#1}}}}  	%roman superscript
\newcommand{\diff}{\ensuremath{{\mathrm{d}}}}		%roman differential operator d

\begingroup
  \catcode`\_=\active
  \gdef_#1{\ensuremath{\sb{\mathrm{#1}}}}	%old _ is now \sb
\endgroup
\mathcode`\_=\string"8000
\catcode`\_=12

\DeclareMathOperator{\dex}{dex}

\begin{document}

  \title{Effects of turbulence and rotation on protostar formation as a precursor of massive black holes}
  %\titlerunning{Effects of turbulence and rotation on protostar formation}

  \author{C.\ Van Borm\inst{\ref{inst1}\fnmsep\ref{inst2}}\fnmsep\thanks{Corresponding author, \email{borm@astro.rug.nl}}
	\and S.\ Bovino\inst{\ref{inst1}} 
	\and M.\ A.\ Latif\inst{\ref{inst1}} 
	\and D.\ R.\ G.\ Schleicher\inst{\ref{inst1}}
	\and M.\ Spaans\inst{\ref{inst2}} 
	\and T.\ Grassi\inst{\ref{inst3}\fnmsep\ref{inst4}}
	}
	
  \institute{Institut für Astrophysik, Georg-August-Universität Göttingen, Friedrich-Hund-Platz 1, 37077 Göttingen, Germany\label{inst1}
  \and
  Kapteyn Astronomical Institute, University of Groningen, PO Box 800, 9700 AV Groningen, The Netherlands\label{inst2}
  \and
  Centre for Star and Planet Formation, Natural History Museum of Denmark, Øster Voldgade 5-7, 1350 Copenhagen, Denmark\label{inst3}
  \and
  Niels Bohr Institute, University of Copenhagen, Juliane Maries Vej 30, 2100 Copenhagen, Denmark\label{inst4}
  }

  \date{Received <date> / Accepted <date>}

  \abstract
  %Context
  {The seeds of the first supermassive black holes may have resulted from the direct collapse of hot primordial gas in $\gtrsim \SI{e4}{K}$ haloes, forming  a supermassive or quasistar as an intermediate stage.}
  %Aims
  {We explore the formation of a protostar resulting from the collapse of primordial gas in the presence of a strong Lyman-Werner radiation background. Particularly, we investigate the impact of turbulence and rotation on the fragmentation behaviour of the gas cloud. We accomplish this goal by varying the initial turbulent and rotational velocities.} 
  %Methods
  {We performed 3D adaptive mesh refinement simulations with a resolution of 64 cells per Jeans length using the ENZO code, simulating the formation of a protostar up to unprecedentedly high central densities of \SI{e21}{cm^{-3}} and spatial scales of a few solar radii. To achieve this goal, we employed the KROME package to improve modelling of the chemical and thermal processes.} 
  %Results
  {We find that the physical properties of the simulated gas clouds become similar on small scales, irrespective of the initial amount of turbulence and rotation.
  After the highest level of refinement was reached, the simulations have been evolved for an additional \about 5 freefall times.
  A single bound clump with a radius of \SI{2e-2}{AU} and a mass of \about \SI{7e-2}{\Msun} is formed at the end of each simulation, marking the onset of protostar formation. 
  No strong fragmentation is observed by the end of the simulations, regardless of the initial amount of turbulence or rotation, and high accretion rates of a few solar masses per year are found.}
  %Conclusion
  {Given such high accretion rates, a quasistar of \SI{e5}{\Msun} is expected to form within \SI{e5}{years}.}


\maketitle

\section{Introduction}
Several very bright quasars have been detected at $z > 6$, which suggests that supermassive black holes (SMBHs) with masses of $\about \SI{e9}{\Msun}$ already existed when the Universe was less than \SI{1}{Gyr} old \citep{2006NewAR..50..665F,2010AJ....139..906W,2011Natur.474..616M,2013ApJ...779...24V}. It is challenging to explain how such SMBHs could have assembled so soon after the Big Bang, in particular how and when the `seeds' of these SMBHs formed and how their subsequent growth proceeded. 
Various scenarios for the formation of seed black holes in the early Universe have been proposed and are briefly discussed below (for a detailed review, see \citet{2010A&ARv..18..279V,2013ASSL..396..293H}).

Perhaps the most obvious scenario assumes that SMBHs grow from the first stellar remnants. The first stars are thought to form at redshifts of $\about \numrange{20}{50}$ in minihaloes of $\about \SI{e6}{\Msun}$, cooled by molecular hydrogen \citep{1997ApJ...474....1T,2002Sci...295...93A,2002ApJ...564...23B,2013ApJ...772L...3L,2013MNRAS.434L..36B,2014ApJ...781...60H}. 
The first generation of stars was expected to have a more top-heavy initial mass function (typical stellar masses \about \SI{100}{\Msun}) than the current mode of star formation, resulting from the inefficient cooling in these minihaloes. However, more recent simulations that follow the collapse beyond the formation of the first core find that fragmentation may be effective after all, and thus the first stars may tend to form in clusters with much lower individual masses than initially expected, $\lesssim$ \SI{10}{\Msun} \citep{2009Sci...325..601T,2010MNRAS.403...45S,2012MNRAS.422..290S,2011Sci...331.1040C,2011ApJ...737...75G,2014MNRAS.441.2181B,2014ApJ...792...32S}. 
Accretion luminosity does not seem to have much influence on the fragmentation behaviour \citep{2011MNRAS.414.3633S,2012MNRAS.424..457S}. On the other hand, stellar UV feedback appears to inhibit accretion onto the protostar, which would result in an upper limit on the stellar mass of $\about \SIrange{50}{100}{\Msun}$ \citep{2011Sci...334.1250H,2012ApJ...760L..37H,2013ApJ...773..185S}. 
Even if very massive stars were able to form and collapse into seed black holes, it would be difficult for them to accrete sufficient mass in the available time. It has been suggested that super-Eddington accretion may be necessary to accomplish this \citep[e.g.][]{2014ApJ...784L..38M}. However, the HII region formed around the seed black hole significantly reduces the accretion rate onto the seed \citep{2009ApJ...696L.146M,2009ApJ...698..766M,2009ApJ...701L.133A,2011MNRAS.410..919J}. 
In addition, it has been found that stars with masses \about \SIrange{100}{10000}{\Msun} may form in massive primordial haloes irradiated by a moderate UV background, with a strong correlation between the strength of the UV flux and the mass of a protostar \citep{2014ApJ...792...78L}.

Another scenario predicts the formation of seed black holes from very compact nuclear star clusters, which may form at redshifts of $\about \numrange{10}{15}$, after some metal enrichment has occurred so that metal line-cooling becomes effective, and in the presence of trace amounts of dust \citep{2003Natur.422..869S,2004Natur.428..724P,2012MNRAS.419.1566S,2012MNRAS.423L..60S,2005ApJ...626..627O,2012PASJ...64..114O,2008ApJ...672..757C,2012MNRAS.421.3217K,2011ApJ...729L...3D,2013ApJ...766..103D,2014MNRAS.438.1669S,2014ApJ...790L..35B}. In such a cluster, stellar collisions can occur in a runaway fashion and lead to the formation of a very massive star, finally resulting in a seed black hole with a mass up to $\about \SI{3000}{\Msun}$ \citep{1978MNRAS.185..847B,2009ApJ...694..302D,2010MNRAS.409.1057D,2012MNRAS.421.1465D,2014MNRAS.442.3616L}. Although this is significantly more massive than what is expected for the first generation of stars, it will still be difficult for such seeds to grow into the observed SMBHs in the available time. 

In this work, we focus on a third pathway: the so-called direct collapse scenario. In this case, the primordial gas in a halo would collapse directly into a single central object, without fragmenting \citep[e.g.][]{2003ApJ...596...34B,2004MNRAS.354..292K,2006MNRAS.370..289B,2006MNRAS.371.1813L,2006ApJ...652..902S,2010ApJ...712L..69S,2013MNRAS.433.1607L,2014MNRAS.439.1160R}. The most likely host candidates are haloes with virial temperatures $\gtrsim \SI{e4}{K}$ at redshifts $\about \numrange{10}{15}$. For a direct collapse to occur, it is important that fragmentation is suppressed, which is possible if the gas in the halo is kept hot (and thus the Jeans mass high). Hence, the formation of \ce{H2} must be inhibited so cooling can occur only through atomic hydrogen, because otherwise molecular hydrogen cooling will lower temperatures to $\about \SI{200}{K}$ and fragmentation may occur. In the absence of \ce{H2} cooling, self-gravitating gas will collapse nearly isothermally until it becomes optically thick and the adiabatic phase sets in. 

One plausible mechanism for suppressing the formation of sufficient \ce{H2} is the presence of a UV radiation background.  
If massive, the first generation of stars (Pop III) is expected to have a stellar spectrum with a characteristic temperature of \about \SI{e5}{K} \citep[T5 spectrum;][]{2000ApJ...528L..65T,2001ApJ...552..464B,2002A&A...382...28S}, while the lower mass second generation of stars  (Pop II) has a softer spectrum with several \SI{e4}{K} (T4 spectrum). These two spectral types have been used in several studies \citep[e.g.][]{2001ApJ...546..635O,2003ApJ...596...34B,2010MNRAS.402.1249S}.
Lyman-Werner radiation $\left(\SIrange{11.2}{13.6}{eV}\right)$ with an intensity above a certain threshold is able to photo-dissociate \ce{H2} and \ce{H-} (important for the formation of \ce{H2}) and keep their abundance very low. A T5 spectrum will mainly directly photo-dissociate \ce{H2}, while a T4 spectrum will be better at photo-detaching \ce{H-}. 
The critical intensity required to suppress \ce{H2} formation in massive haloes where direct gas collapse can occur has been estimated at $J_{21}\ssup{crit} \gtrsim \numrange{e2}{e3}$, where $J_{21}$ denotes the specific intensity just below the Lyman limit (\SI{13.6}{eV}), in units of \SI{e-21}{erg. cm^{-2}.sr^{-1}.s^{-1}.Hz^{-1}} \citep[e.g.][]{2001ApJ...546..635O, 2003ApJ...596...34B, 2010ApJ...712L..69S, 2010MNRAS.402.1249S, 2013A&A...553L...9V, 2014MNRAS.443.1979L,2014ApJ...792...78L}. This is relatively high compared to the expected cosmic UV background at the relevant redshifts ($J_{21}^{\mathrm{bg}} \sim 10$ at $z \sim 10$) \citep[e.g.][]{2009ApJ...695.1430A, 2012MNRAS.419..718H, 2014MNRAS.442.2036D}.
However, the UV background distribution has a long bright-end tail, owing to the presence of close (about \SI{10}{kpc}) luminous neighbours, which means that there is a small but significant subset of haloes that is exposed to supercritical intensities \citep{2008MNRAS.391.1961D,2014MNRAS.443..648A,2014MNRAS.442.2036D,2014arXiv1406.7020V}. 
Recently, though, it has been shown that it is important to consider spectra generated from realistic stellar populations, taking the mode of star formation (continuous or bursty) and the age, metallicity, and mass of the stars into account \citep{2014arXiv1407.4115A,2014arXiv1407.4039S}. This has implications for the \ce{H2} photo-dissociation rate and the \ce{H-} photo-detachment rate, and thus affects the value of the critical intensity, $J_{21}\ssup{crit}$.
\citep{2014arXiv1407.4115A} computed the reaction rate coefficients for \ce{H2} photo-dissociation and \ce{H-} photo-detachment using realistic spectra resulting from a stellar synthesis code, and found that these depend on the age and metallicity of the stars, in contrast to the findings of \citep{2014arXiv1407.4039S}. The latter used a one-zone model and realistic stellar spectra to also calculate $J_{21}\ssup{crit}$, finding values in the range between \num{1000} and \num{1400}.
\citet{2014arXiv1408.3061L} have studied the impact of varying the temperature of a blackbody spectrum in 3D cosmological simulations, to more closely resemble a realistic spectrum generated by Pop II stars. They found an even higher value for $J_{21}\ssup{crit}$, a few times \num{e4}, due to additional 3D effects. This value depends only weakly on the adopted radiation spectra in the range between $T_{rad} = \SI{2e4}{K}$ and \SI{e5}{K}.

Alternative mechanisms for inhibiting \ce{H2} cooling comprise dissipation of a sufficiently strong magnetic field \citep{2009ApJ...703.1096S,2010ApJ...721..615S,2013A&A...553L...9V} or the presence of strong shocks \citep{2012MNRAS.422.2539I}, both of which result in collisional dissociation of \ce{H2}.

Numerical 3D simulations have found fragmentation to be inhibited and thus show the feasibility of the direct collapse scenario. In some simulations, bar-like instabilities \citep{2008ApJ...682..745W} or self-gravitating disks on parsec scales \citep{2009MNRAS.396..343R} were found, though these employed a resolution of 16 cells per Jeans length. More recently, it was demonstrated that at least 32 cells, and preferably more, are required to properly resolve turbulence \citep{2010ApJ...721L.134S,2011ApJ...731...62F,2012ApJ...745..154T,2013MNRAS.430..588L}. New simulations employing a higher resolution find that it is likely that \about \SI{e5}{\Msun} objects will form \citep{2013MNRAS.433.1607L,2013MNRAS.430..588L,2013MNRAS.436.2989L}, though the peak density in these studies is not much higher than \SI{e15}{cm^{-3}}. In the simulations pursued here, we aim to complement these studies exploring collapse and fragmentation on smaller scales.

While simulating the formation of the protostar in 3D was not yet possible, various one-zone models employing detailed chemical models show the expected thermal pathway \citep{2000ApJ...534..809O,2001ApJ...546..635O,2005ApJ...626..627O,2008ApJ...686..801O}. For a strong UV background, \citet{2001ApJ...546..635O} showed that clouds collapse nearly isothermally, cooled successively by Lyman-alpha emission of atomic hydrogen, two-photon emission of atomic hydrogen from the $2s$ state, and \ce{H-} free-bound emission. Afterwards, the adiabatic phase sets in at \about \SI{e20}{cm^{-3}}, at which point the minimum Jeans mass, and thus the characteristic mass of the protostar, has been reduced to \SI{0.03}{\Msun}.

Once the protostar has formed, it will accrete and evolve into either a supermassive star or a quasistar, depending on the accretion rate. The work by \citet{2013A&A...558A..59S} suggests that for accretion rates >\SI{0.14}{\Msun/yr}, a quasistar will be the result, while lower accretion rates lead to the formation of a supermassive star.
A supermassive star (SMS, with a mass in the range \SIrange{e3}{e6}{\Msun}) of fixed mass, supported by radiation pressure, is thought to evolve as an $n = 3$ polytrope and finally collapse into a black hole containing most of the stellar mass \citep{2011MNRAS.410..919J,2013ApJ...774...64W,2012ApJ...756...93H,2013ApJ...778..178H}. 
However, if the mass accretion rate is high enough, the outer layers of the SMS cannot thermally relax. In this case, it is not well-described by an $n = 3$ polytrope, but will have a more complex structure with a convective core surrounded by a convectively stable envelope that contains most of the mass. The core will burn up its hydrogen, and subsequently collapse into a black hole with a mass of a few \Msun. The resulting structure, where the black hole accretes material from the massive, radiation-pressure-supported envelope, is termed a `quasistar' \citep{2006MNRAS.370..289B,2008MNRAS.387.1649B,2010MNRAS.402..673B,2010MNRAS.409.1022V,2011MNRAS.414.2751B,2012MNRAS.421.2713B}.

As is known from present-day star formation, turbulence plays an important role in angular momentum transport and determining the fragmentation properties of collapsing gas clouds, since it can both locally compress the gas as well as provide additional support against collapse on larger scales \citep[e.g.][]{1981MNRAS.194..809L,2004RvMP...76..125M,2007ARA&A..45..565M,2012ApJ...761..156F}. Similar effects have been found at high redshifts in simulations of minihaloes, where turbulence plays a role in distributing angular momentum \citep{2002Sci...295...93A}, and affects the fragmentation behaviour \citep{2011ApJ...727..110C,2012ApJ...745..154T,2013ApJ...772L...3L}. Also in simulations of more massive, atomic cooling haloes, the importance of turbulence has been recognized \citep{2008MNRAS.387.1021G,2008ApJ...682..745W}. However, many of these older studies do not employ a sufficient Jeans resolution, as its impact was only recognized later. \citet{2013MNRAS.430..588L} found that the amount of turbulent structure increases significantly with increasing resolution, and in the study by \citet{2013MNRAS.433.1607L} it was found that fragmentation occurs occasionally, but that this does not prevent the growth of a central massive object resulting from turbulent accretion and mergers.

Numerical simulations of collapsing gas in minihaloes show that fragmentation also depends on the amount of rotation, with stronger rotation inducing more fragmentation \citep{2002ApJ...564...23B,2008ApJ...682L...1M,2010A&A...510A.110H}. The study by \citet{2008ApJ...672..757C} shows that massive disk-like structures are assembled, fragmenting to form protostars.  In atomic cooling haloes the effects of rotation have not yet been studied in detail, though \citet{2003ApJ...596...34B} found that a single black hole is formed in low-spin galaxies, while higher spin galaxies tend to form binary black holes. 
In their simulations of atomic cooling haloes, \citet{2009MNRAS.393..858R} observed the formation of massive compact self-gravitating disks, and found mild fragmentation in one of the three simulated haloes.

In this paper we present the first study in which the formation of a massive protostar is simulated in 3D up to unprecedented high central densities (\SI{e21}{cm^{-3}}), owing to improved modelling of the chemistry. A high spatial resolution is obtained as well; starting from pc scales, we are able to resolve scales down to a few solar radii. In addition, we investigate how the fragmentation behaviour of collapsing primordial gas in the presence of a strong Lyman-Werner radiation background is affected by varying amounts of turbulence and rotation. For each case the formation of clumps and their accretion rates are studied. 

In Sect.~\ref{sec:methods} some details are given on the methods and setup of the numerical simulations that have been performed. In Sect.~\ref{sec:results} the results for both the one-zone calculations and the 3D simulations are presented and discussed, and we conclude with a summary of the results in Sect.~\ref{sec:conclusion}.

\section{Numerical Methodology and Simulation Setup} \label{sec:methods}
ENZO is an open-source adaptive mesh refinement (AMR) simulation code, which provides high spatial and temporal resolution for the modelling of astrophysical fluid flows \citep{2014ApJS..211...19B}. 
It contains a wide variety of physics modules, making it suitable for many different astrophysical applications. We use a modified version of ENZO 2.3, replacing the chemistry implementation by a customized build of the KROME chemistry package \citep{2014MNRAS.tmp..321G}, as discussed in the following subsections. 
The hydrodynamical equations are solved using the MUSCL scheme, which is a second-order accurate extension of Godunov's method.
The implementation in Enzo uses second-order Runge-Kutta time integration, and the Riemann solver employed is the HLLC solver (Harten-Lax-van Leer with Contact), with a fallback to the more diffusive HLL solver (Harten-Lax-van Leer) in case negative energies or densities are computed. 
The choice of this solver is due to the strong shocks which occur once the central core becomes adiabatic and the central protostars forms. 
Self-gravity is computed by solving the Poisson equation using a multigrid technique.

\subsection{Initial conditions}
We follow the gravitational collapse of an isolated spherical gas cloud with a radius of \SI{15}{pc} and a top-hat density profile, embedded in a \SI{100}{pc} simulation box. 
The Jeans length is resolved by at least 64 cells at all times. Additionally, a refinement criterion based on overdensity is used. 
These combined criteria result in the simulations using 29 refinement levels, at which point an adiabatic core is formed and no further refinement is necessary. The collapse is followed for another \SI{1.67e-2}{years}, corresponding to \about 5 freefall times, after the highest refinement level is reached. 
To ensure pressure equilibrium between the sphere and its surroundings, we set the initial sphere density to \SI{1000}{cm^{-3}} and its temperature to \SI{500}{K}, while the surrounding gas is initialized with a density of \SI{100}{cm^{-3}} and a temperature of \SI{5000}{K}. The above combination of parameters also ensures that the mass of the cloud (\about \SI{3.5e5}{\Msun}) is greater than the local Jeans mass (\about \SI{3e4}{\Msun}), and thus the cloud will collapse. The total mass contained in the box is \about \SI{2.8e6}{\Msun}. This setup has been chosen in order to be able to explore the formation of  protostars up to very high densities.
The cloud is irradiated by a UV background with a T5 spectrum (see \ref{subsubsec:radbg}) of intensity \num{e5} in units of $J_{21}$,  
so that the abundance of \ce{H2} is kept low and cooling will occur mainly through atomic hydrogen. 

Furthermore, we add a certain amount of initial turbulence to the gas, as well as some rotation of the cloud. These parameters are varied to study and quantify their effects on the collapse dynamics and fragmentation properties. An overview of the different simulations can be found in Table~\ref{tab:simpars}. The turbulent velocities are sampled from a Maxwellian distribution with a temperature equal to the initial temperature of the gas sphere, and subsequently multiplied by the percentage given in the table. Since the maximum of the Maxwell distribution function is of the order of the sound speed $c_s$, the turbulent velocities are of the order of a given percentage of $c_s$.
The amount of rotation is given in percentage of the Keplerian velocity; i.e.\ 100\% rotation means the cloud is rotationally supported.

\begin{table}[!htbp]
\caption{Overview of the different simulations and their initial turbulent and rotational velocities.}
\label{tab:simpars}
\centering
\begin{tabular}{lcc} 
	\hline\hline
	\multicolumn{3}{c}{\textbf{Simulations}} \\[3pt]	
	Name & Turbulence ($\about$ \% of $c_s$) & Rotation (\% of $v_{Kep}$) \\ 
	\hline
	T40R0  & 40 \% & 0  \% \\
	T40R10 & 40 \% & 10 \% \\
	T40R20 & 40 \% & 20 \% \\
	T20R10 & 20 \% & 10 \% \\
	T80R10 & 80 \% & 10 \% \\
	\hline
\end{tabular}
\tablefoot{Turbulent velocities are sampled from a Maxwellian distribution where the temperature is the initial temperature of the gas sphere, and subsequently multiplied by the percentage given in the table. The amount of rotation is given in percentage of the Keplerian velocity.}
\end{table}

\subsection{Chemistry, heating, and cooling} \label{subsec:chemistry}
We employ the KROME\footnote{Publicly available at \url{http://kromepackage.org/}} chemistry package, which has been developed in order to simplify the embedding of the chemistry and the microphysics in numerical simulations. 
It builds the corresponding rate equations, the solver parameters, and includes a series of thermal processes which are coupled to the chemical evolution. 
A patch to embed KROME in ENZO is available with the package and has been used within this work.
KROME solves the non-equilibrium chemistry together with the temperature equation using the adaptive high-order solver DLSODES, which was shown to be both accurate and efficient for networks that present a corresponding ordinary differential equation system with a sparse Jacobian, and that are typical in astrophysical applications \citep{2013MNRAS.434L..36B,2013MNRAS.431.1659G}.
We have modified and extended the available package, mainly to obtain the desired cooling processes.
The main improvements of our modified version of KROME are the addition of \ce{H-} cooling, Rayleigh scattering, and a different evaluation of the critical density used for the chemical heating, following \citet{2008MNRAS.388.1627G}.

\subsubsection{Chemical network}
Our chemical network consists of 31 reactions, including 9 species: \ce{H}, \ce{e}, \ce{H+}, \ce{H-}, \ce{H2}, \ce{H2+}, \ce{He}, \ce{He+}, and \ce{He++}. All the reactions and their associated rates can be found in Appendix~\ref{app:rates}.

\subsubsection{Molecular hydrogen cooling}
The molecular hydrogen cooling rates were taken from \citet{2008MNRAS.388.1627G}, with an opacity correction from \citet{2004MNRAS.348.1019R}, as implemented in KROME. However, we modified the opacity correction to use the molecular hydrogen density instead of the total density, rendering it usable for cases with a non-zero UV background. Hence, the \ce{H2} cooling rate is multiplied by a factor $\min{\left[1, \left(n_{H2} / \left(\SI{8e9}{cm^{-3}}\right)\right)^{-0.45}\right]}$, where $n_{H2}$ is the \ce{H2} number density. 
Recent studies by \citet{2014MNRAS.444.1566G} and \citet{2014arXiv1407.2102H} calculate the escape fraction of cooling photons using a multi-line, multi-frequency ray-tracing scheme, and an approach based on the TreeCol algorithm, respectively. \citet{2014MNRAS.444.1566G} find that the radially averaged escape fraction agrees well with the analytical fit from \citet{2004MNRAS.348.1019R}, while the results of \citet{2014arXiv1407.2102H} suggest that this fit underestimates the escape fraction after the initial stage of collapse. Presently, it has not yet been investigated which of these two methods yields the most accurate results. However, additional one-zone calculations suggest that even a significantly larger escape fraction does not influence our results, as the ineffectiveness of the cooling is mainly the result of the low \ce{H2} abundance.
Of course, opacity effects would become more important when considering a case where \ce{H2} is the dominant coolant.

\subsubsection{Collision-induced emission cooling}
When a collision takes place between an \ce{H2} molecule and another \ce{H2} molecule, a \ce{He} molecule, or a \ce{H} atom, the interacting pair briefly acts as a `supermolecule' with a non-zero electric dipole, and there is a high probability of a photon being emitted. Collision-induced emission (CIE) may become important at high densities, depending on the gas temperature.
We use the fit provided in KROME for the optically thin rate, but modified to ensure it is 0 if $f_{\ce{H2}} < 0.1$ and does not become important before $f_{\ce{H2}} \sim 0.5$, where $f_{\ce{H2}}$ is the \ce{H2} mass fraction relative to \ce{H}, as it is uncertain whether the fit is still valid for extremely dissociated media. The opacity correction at high densities has been adopted from \citet{2004MNRAS.348.1019R},
\begin{equation}
 \tau_{CIE} = \max{\left[\num{e-5}, \left(\frac{n}{\SI{2e16}{cm^{-3}}}\right)^{2.8}\right]},
\end{equation}
where $n$ is the total number density. The CIE cooling rate is then multiplied by $\min{\left[1, \left(1 - \exp{\left(-\tau_{CIE}\right)}\right) / \tau_{CIE}\right]}$, where $\left(1 - \exp{\left(-\tau\right)}\right) / \tau$ is the usual spherical escape probability.

\subsubsection{Atomic cooling} \label{subsubsec:atomcooling}
KROME employs the atomic cooling rates from \citet{1992ApJS...78..341C}. These include the collisional ionization of \ce{H}, \ce{He}, \ce{He+}, and \ce{He}(2s) by electrons, the recombination of \ce{H+}, \ce{He+}, and \ce{He++}, the dielectronic recombination of \ce{He+}, the collisional excitation of \ce{H} (all $n$), \ce{He} ($n=$ 2,3,4 triplets), and \ce{He+} ($n=2$), and bremsstrahlung for all ions. The main cooling channel relevant here is the collisional excitation of \ce{H}. We have added an optical depth approximation for the Rayleigh scattering by \ce{H} atoms, which will suppress this main channel, as
\begin{equation}
 \tau_{Rl} = \sigma_{H,Rl} n_{HI} \frac{\lambda_J}{2},
\end{equation}
where 
$\lambda_J$ is the Jeans length, $n_{HI}$ is the number density of atomic hydrogen, and 
\begin{align}
   \sigma_{H,Rl} &= \num{5.799e-29} \lambda^{-4} + \num{1.422e-30} \lambda^{-6} \notag\\
		 &\quad\quad+ \num{2.784e-32} \lambda^{-8} \si{cm^2}
\end{align}
is the Rayleigh scattering cross section of \ce{H} for radiation with wavelength $\lambda$ (in \si{$\mu$m}) \citep{1970SAOSR.309.....K}. The cooling rate is then multiplied by $\exp{\left(-\tau_{Rl}\right)}$.
Additionally, we have added two fudge factors to mimic optical depth effects and thus reduce cooling at high densities ($n \gtrsim \SI{e17}{cm^{-3}}$), in accordance with the findings of \citet{2001ApJ...546..635O}. 
The first factor, $f_1$, represents that the gas should be optically thick to atomic hydrogen line cooling around \about \SI{e17}{cm^{-3}}, and \ce{H} ionization becomes the main atomic cooling channel. The second factor, $f_2$, ensures that the gas becomes almost completely optically thick to radiative cooling around \about \SI{e20}{cm^{-3}}, so that afterwards the evolution is nearly adiabatic.
The fudge factors are calculated as $f_i = \min{\left[1, \left(1 - \exp{\left(-\tau_{f_i}\right)}\right) / \tau_{f_i}\right]}$ for $i=1,2$, using the functional form of the spherical escape probability. The opacities $\tau_{f_1}$ and $\tau_{f_2}$ are given by
\footnote{The exponents of 5 and 8 do not have a specific physical meaning, but are instead intended to provide a sharp enough cutoff, as in this regime the atomic cooling functions increase steeply with both density and temperature.}
\begin{align}
 \tau_{f_1} &= \max{\left[\num{e-5}, \left(\frac{n}{\SI{e17}{cm^{-3}}}\right)^5\right]}, \\
 \tau_{f_2} &= \max{\left[\num{e-5}, \left(\frac{n}{\SI{e20}{cm^{-3}}}\right)^8\right]}.
\end{align}

\subsubsection{\texorpdfstring{\ce{H-} cooling}{H- cooling}}
Through radiative association of \ce{H} and \ce{e}, \ce{H-} is formed and a photon is emitted. There will be net cooling if this photon can escape \citep{2001ApJ...546..635O,2008A&A...490..521S}. The cooling rate can then be approximated as
\begin{equation}
 \Lambda_{\ce{H-}} \approx k_{\ce{H-}} n_{HI} n_{e} E_{\gamma},
\end{equation}
where $E_{\gamma}$ is the approximate energy of the emitted photon. A typical electron undergoing radiative attachment has an energy of the order of $k_B T$, so the average outgoing photon energy can be estimated as $E_{\gamma} \sim E_0 + k_B T$, where the binding energy $E_0$ of \ce{H-} is \SI{0.755}{eV}. Rayleigh scattering (see Sect.~\ref{subsubsec:atomcooling}), as well as \ce{H-} bound-free absorption, will suppress this cooling channel, so optical depth approximations for these processes have been taken into account. The cross section for \ce{H-} bound-free absorption is \citep{1988A&A...193..189J}
\begin{equation}
 \sigma_{\ce{H-},bf} = \num{e-18} \lambda^3 \left(\frac{1}{\lambda} - \frac{1}{\lambda_0}\right)^{1.5} f\left(\lambda\right),
\end{equation}
where $\lambda$ is the wavelength of the scattered radiation in \si{$\mu$m}, $\lambda_0 = \SI{1.6419}{$\mu$m}$, and $f\left(\lambda\right)$ is given by equation~5 in \citet{1988A&A...193..189J}.

\subsubsection{Chemical cooling and heating} \label{subsubsec:chemhc}
Various chemical reactions can result in net cooling or heating of the gas \citep{2000ApJ...534..809O}. In our case, the most important ones are the three-body formation of \ce{H2} (\citet{2013ApJ...773L..25F}, see \citet{2014A&A...561A..13B} for a comparison of different rates) and collisional dissociation of \ce{H2} (\citet{1987ApJ...318...32S,1996ApJ...461..265M,1998ApJ...499..793M}). The collisional dissociation process releases \SI{4.48}{eV} per dissociated \ce{H2} molecule (its binding energy), cooling the gas, while the heat deposited by three-body formation is $4.48 (1+n_{cr}/n)^{-1}\si{eV}$ per \ce{H2} molecule. Here, $n_{cr}$ is the critical density, calculated as \citep{2008MNRAS.388.1627G}
\begin{equation}
 n_{cr} = \left(\frac{x_{\ce{HI}}}{n_{cr,\ce{HI}}} + \frac{x_{\ce{H2}}}{n_{cr,\ce{H2}}}\right)^{-1},
\end{equation}
where $x_{\ce{HI}}$ and $x_{\ce{H2}}$ are the number fractions of \ce{HI} and \ce{H2}, respectively, and $n_{cr,\ce{HI}}$ and $n_{cr,\ce{H2}}$ are their respective critical densities, given by
\begin{align}                                                                        
 n_{cr,\ce{HI}} &= \dex{\left[3 - 0.416 \log{T_4} - 0.327 \left(\log{T_4}\right)^2\right]}, \\
 n_{cr,\ce{H2}} &= \dex{\left[4.845 - 1.3 \log{T_4} + 1.62 \left(\log{T_4}\right)^2\right]},
\end{align}
where $T_4 = \frac{T}{\SI{e4}{K}}$.

\subsubsection{Radiation background} \label{subsubsec:radbg}
In our calculations we have used a constant UV background flux with a T5 spectrum below the Lyman limit, which will photo-dissociate \ce{H2} and photo-detach \ce{H-}. The main difference with a T4 spectrum is that lower values of the intensity, $J_{21}$, are required for the gas to collapse isothermally.
We do not expect the choice of spectrum or the specific strength of the UV background to matter, as long as the \ce{H2} abundance is kept low so that \ce{H2} cooling is unimportant.
The difference between the spectra is expressed in the photo-dissociation rate of \ce{H2} and photo-detachment rate of \ce{H-} (see $k_{24}$ and $k_{25}$ in Appendix~\ref{app:rates}). 
We also include \ce{H2} self-shielding, using the improved fit described in \citet{2011MNRAS.418..838W},
\begin{align}
 f_{sh} &= \frac{0.965}{\left(1 + x_{N_{H_2}}/b_5\right)^{1.1}} + \frac{0.035}{\left(1 + x_{N_{H_2}}\right)^{0.5}} \notag\\
        &\quad\quad\exp{\left(\num{-8.5e-4} \left(1+x_{N_{H_2}}\right)^{0.5}\right)},
\end{align}
where $x_{N_{H_2}}$ is given by
\begin{equation}
 x_{N_{H_2}} = \frac{N_{H_2}}{\SI{5e14}{cm^{-2}}}, 
\end{equation}
with $N_{H_2}$ the column density in \si{cm^{-2}}, calculated as $N_{H_2} = n_{H_2} \lambda_J/2$.
The Doppler broadening parameter for \ce{H2}, $b_5$, is given by
\begin{equation}
 b_5 = \num{e-5} \left(\frac{2 k_B T}{2 m_H}\right)^{0.5},
\end{equation}
in units of \SI{e5}{cm/s}. The photo-dissociation rate of \ce{H2} is multiplied by the self-shielding factor $f_{sh}$.

\section{Results} \label{sec:results}
\subsection{One-zone calculations} \label{subsec:one-zone}
\begin{figure*}
 \centering
   \includegraphics[width=17cm]{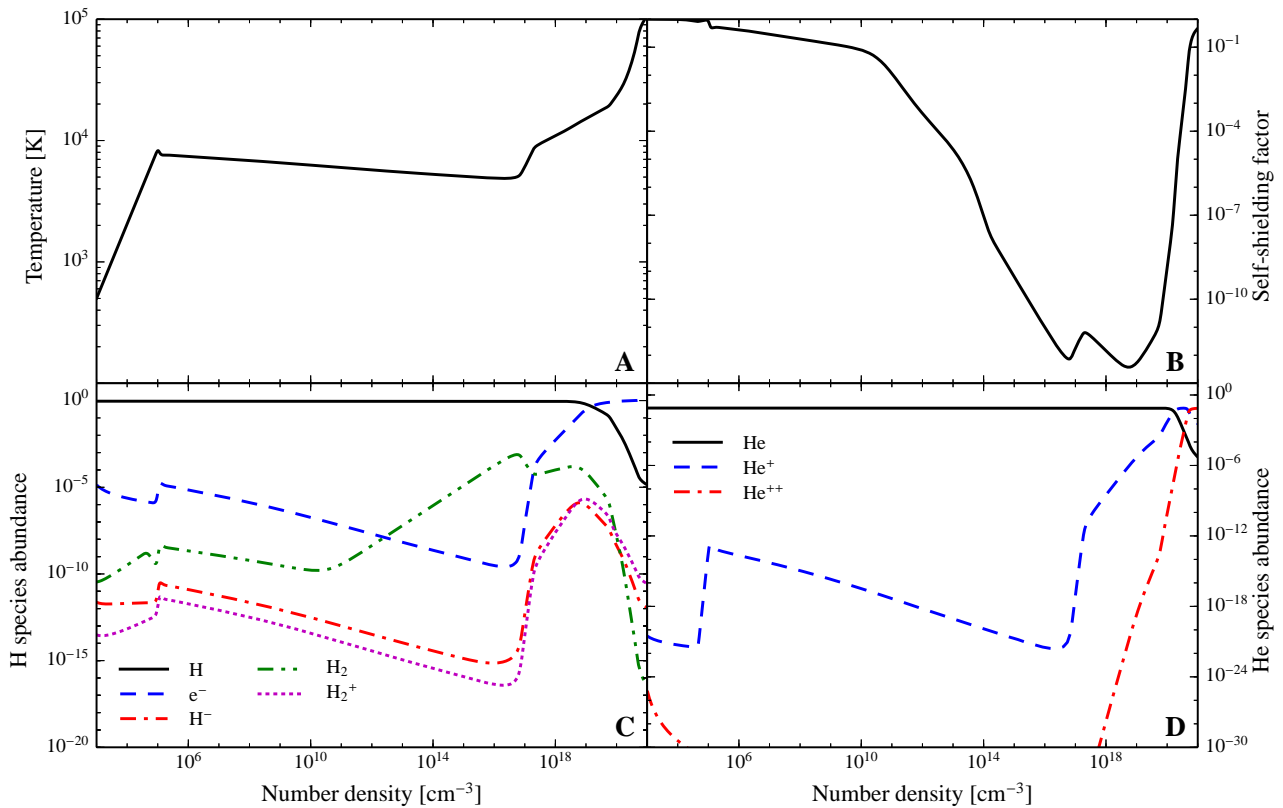}
 \caption{Physical quantities as a function of number density in a one-zone calculation, irradiated by a strong T5 background, using our modification of KROME. A: temperature; B: self-shielding factor for \ce{H2}, $f_{sh}$ ($f_{sh} = 1$ means no shielding, and the smaller $f_{sh}$, the stronger the shielding); C: number fractions of H species (and electrons); D: number fractions of He species.}
\label{fig:one-zone}
\end{figure*}

We have performed a one-zone freefall collapse test, already included with the KROME package, to verify our chemical model. 
The results for a T5 background with $J_{21} = \num{e5}$ and initial conditions similar to those of the 3D simulations can be seen in Fig.~\ref{fig:one-zone}, calculated up to a number density of \SI{e21}{cm^{-3}}. Panel~A shows the temperature evolution, 
panel~B shows the self-shielding factor for molecular hydrogen, panel~C shows the number fractions of the different H species (and electrons), and panel~D shows the number fractions of the different He species.
We note here that at densities above \SI{e17}{cm^{-3}} an equilibrium approximation could be adopted and might speed up the calculations. Nevertheless, we preferred to follow a complete non-equilibrium evolution.

Initially, the temperature increases adiabatically, due to strong compressional heating. Because the molecular hydrogen is strongly dissociated by the UV background, the gas cannot cool through \ce{H2} and instead cools via other processes. During the initial adiabatic phase, \ce{H-} cooling is the dominant cooling process, but it is not efficient enough to counter the strong heating. When the temperature reaches \about \SI{8000}{K}, around \about \SI{e5}{cm^{-3}}, Ly$\alpha$ cooling starts to become dominant and the temperature slope flattens off, now evolving nearly isothermally, though still decreasing slowly. Both chemical cooling and \ce{H-} cooling are also important during this phase. Around a number density of \about \SI{e8}{cm^{-3}}, both of these rates become higher than the atomic cooling. 
The \ce{H-} cooling channel becomes strongly suppressed around \about \SI{e16}{cm^{-3}} as the cloud becomes optically thick to both Rayleigh scattering and \ce{H-} bound-free absorption. Chemical cooling still maintains the near-isothermal evolution briefly, but then chemical heating cancels out the cooling and the temperature starts rising. Collisional ionization of H starts at \about \SI{e17}{cm^{-3}}, resulting in a slowdown of the temperature rise up to \about \SI{e19}{cm^{-3}}. From this point on, the cloud collapses adiabatically, and after sufficient contraction a protostar is expected to form in the centre.

During the whole collapse, the molecular hydrogen fraction never becomes larger than \num{e-3}, and as a result \ce{H2} cooling is not important (except for densities between \SIrange{e4}{e5}{cm^{-3}}). Starting from \about \SI{e10}{cm^{-3}}, three-body formation increases the \ce{H2} abundance, peaking just before the rise in temperature at \about \SI{e17}{cm^{-3}}, after which strong collisional dissociation drastically decreases the abundance again.
At low densities the self-shielding is too weak to prevent \ce{H2} from being photo-dissociated (the smaller the factor, the stronger the shielding). At densities above \SI{e10}{cm^{-3}} the gas starts to become well-shielded, however, due to the high temperature, collisional dissociation of \ce{H2} becomes effective. Additionally, \ce{H2} cooling starts to become optically thick at these densities.

\subsection{3D simulation results}
\begin{figure*}[!htbp]
 \centering
   \includegraphics[width=17cm]{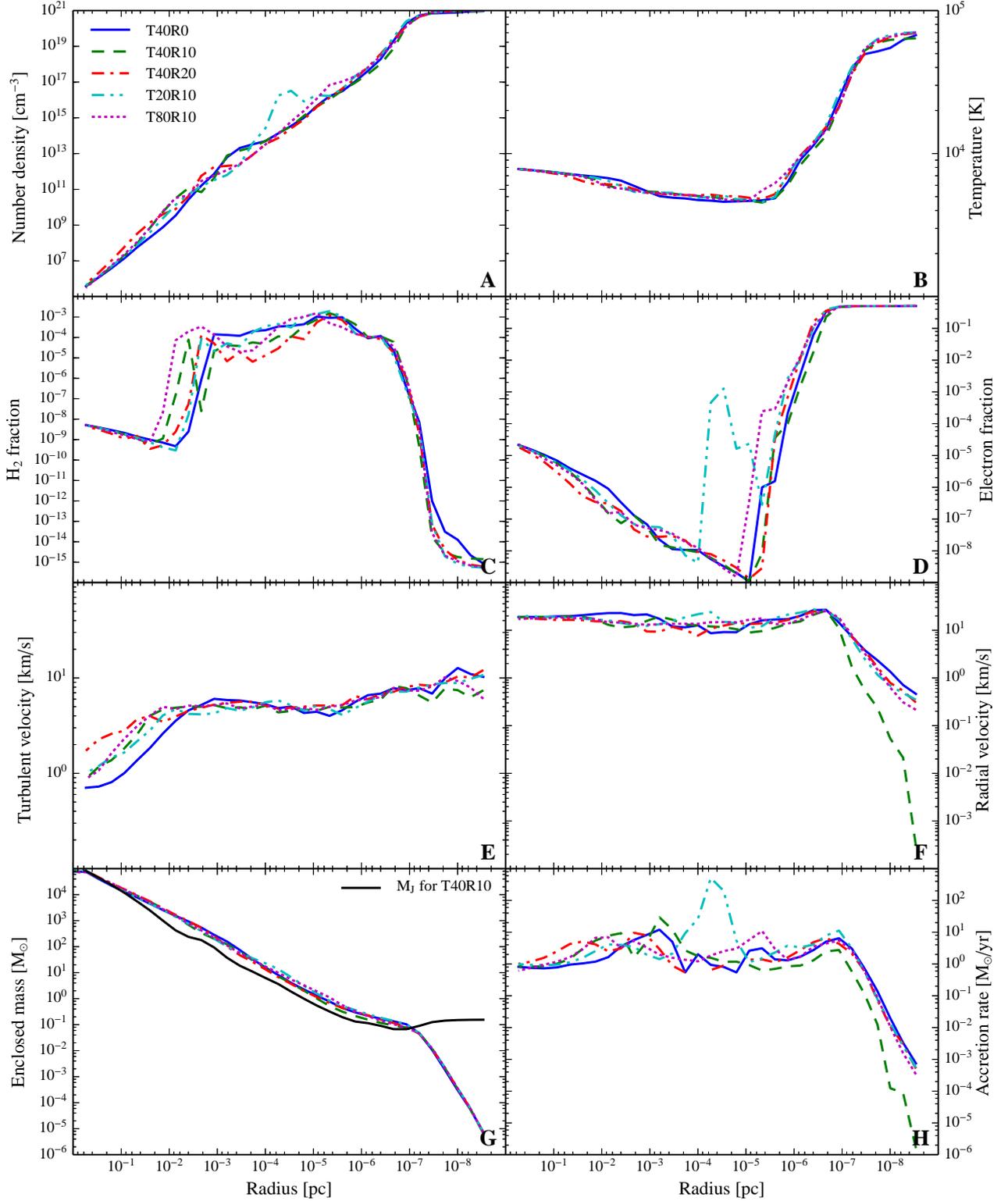}
 \caption{Physical quantities, weighted by mass, spherically averaged and radially binned, as a function of radius at the peak density output for the different simulations. A: number density; B: temperature; C: \ce{H2} number fraction; D: electron number fraction; E: turbulent velocity; F: radial velocity, plotted as $-v_{rad}$; G: enclosed mass, and the Jeans mass for T40R10 (it is very similar for other runs); H: radial mass infall rate, calculated from the density and the radial velocity. The simulation details and abbreviations are listed in Table~\ref{tab:simpars}.}
\label{fig:prfsph}
\end{figure*}

\begin{figure*}[!htbp]
 \centering
   \includegraphics[width=17cm]{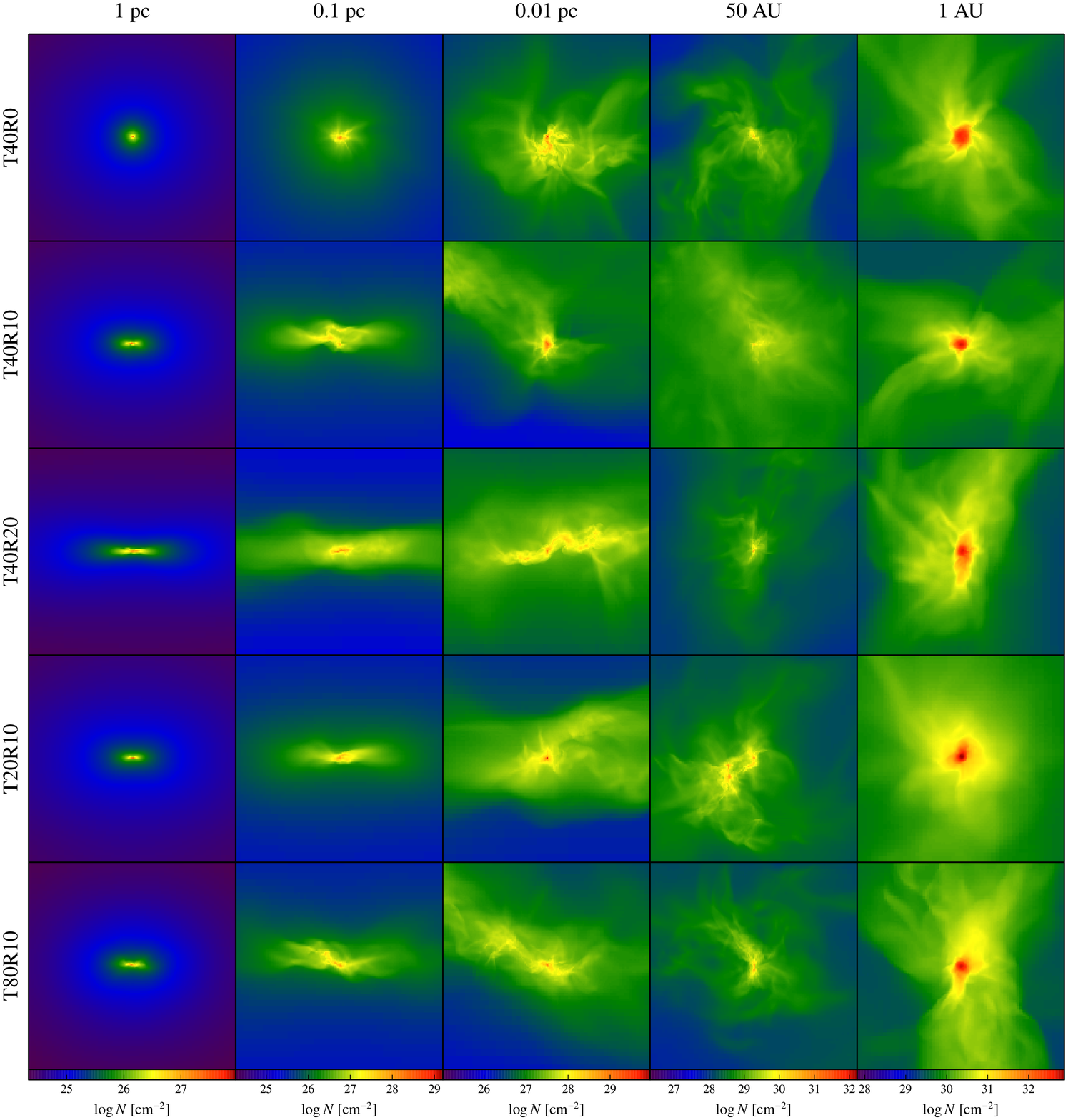}
 \caption{Density projections along the x-axis for all simulations, showing the integrated number density for various scales at the peak density output.The simulation details and abbreviations are listed in Table~\ref{tab:simpars}.}
\label{fig:prjx}
\end{figure*}

\begin{figure*}[!htbp]
 \centering
   \includegraphics[width=17cm]{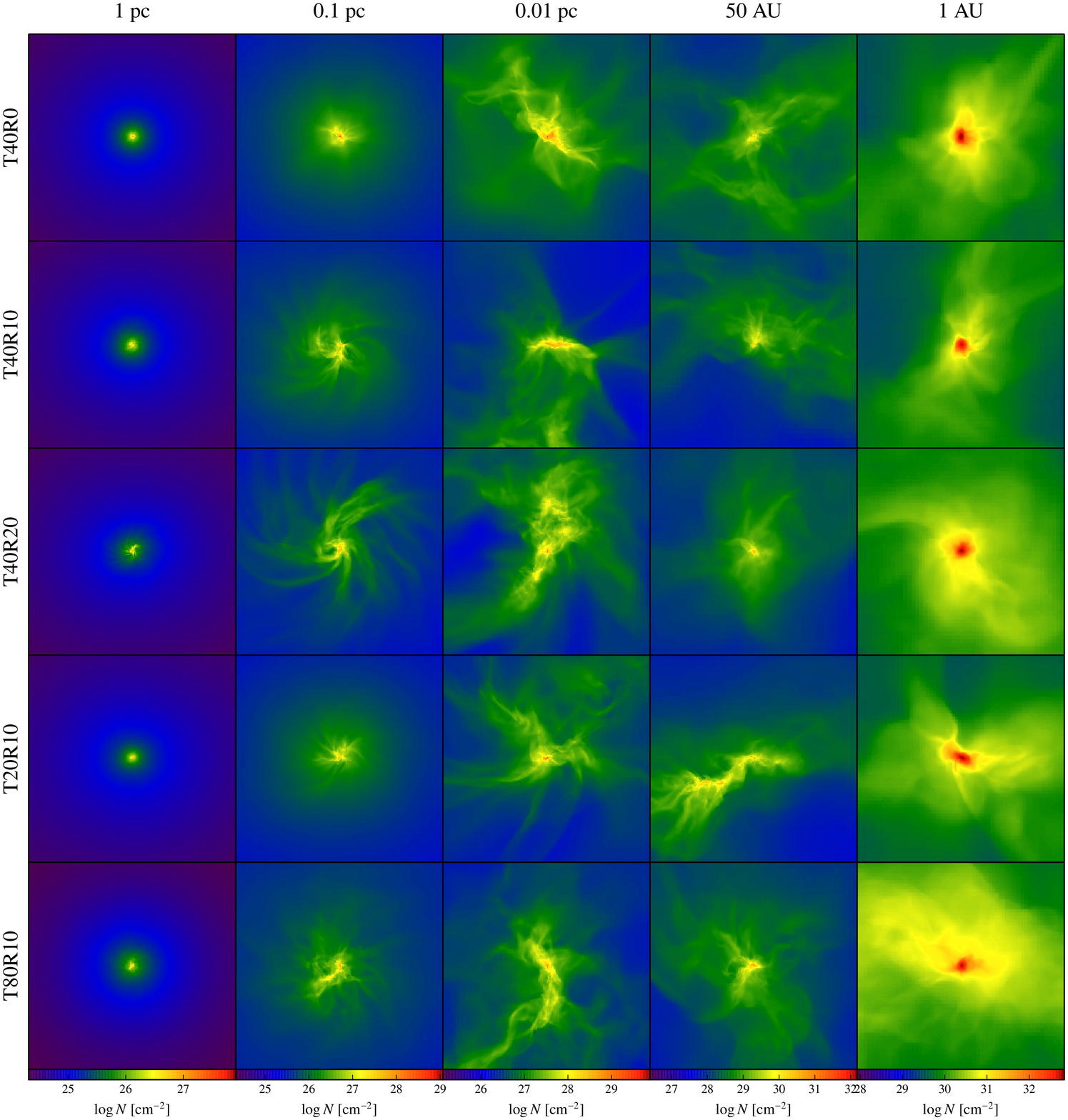}
 \caption{Density projections along the z-axis for all simulations, showing the integrated number density for various scales at the peak density output.The simulation details and abbreviations are listed in Table~\ref{tab:simpars}.}
\label{fig:prjz}
\end{figure*}

An overview of the different 3D simulations and their abbreviations are listed in Table~\ref{tab:simpars}. We have performed one simulation for each set of initial conditions, five in total. After reaching the highest refinement level of 29, the simulations evolved for another \about \SI{1.67e-2}{years}, corresponding to \about 5 freefall times, with the freefall time (\about \SI{3.5e-3}{yr}) calculated at the moment when the highest refinement level is first reached.

Fig.~\ref{fig:prfsph} shows several spherically averaged, radially binned profiles of various quantities for all simulations, centred on the peak density location (hereafter referred to as the central clump). The data shown has been obtained at the end of each simulation, when a peak density of \SI{e21}{cm^{-3}} was reached.
From the density profile (shown in panel~A) it can be seen that in general, the density increases with decreasing radius, so that overall the evolution of quantities with decreasing radius corresponds to an evolution with increasing density. Specifically, the density increases approximately as $\propto r^{-2}$, as is typical of an isothermal collapse. Deviations from this behaviour are caused by local over- or underdensities, resulting from the turbulent nature of the gas.
In the very centre of the cloud, inside of the radius corresponding to the minimum Jeans mass, the density profile flattens off, indicating the central clump.
This clump can also be seen in the enclosed mass profile (shown in panel~G), which steeply decreases inside \SI{e-7}{pc}, due to enhanced pressure support.

The density profile in simulation T20R10 deviates somewhat from isothermal, with a peak in the density profile around \about \SI{4e-5}{pc}. After close inspection of density projections at different scales (see Figs.~\ref{fig:prjx} and \ref{fig:prjz}, particularly at the \SI{50}{pc} scale), this appears to be due to the presence of a second concentration of mass containing two additional clumps, which have not collapsed as far as the main clump.
However, from comparison runs with the same initial conditions, though with a different random seed for the initialization of the turbulent velocity field, we have found that such additional clumps are only sometimes present for the T20R10 initial conditions. Additionally, a second clump is also sometimes found for the other initial conditions discussed, though always with a lower peak density than the main clump. Hence, this fragmentation is likely not related to the amount of initial turbulence or rotation. 
It is not yet clear whether these additional clumps will continue to collapse, or instead accrete onto the main clump.
Based on a simplified ``toy'' model of fragmentation in the accretion disk around a protostar, \citet{2014arXiv1406.5058I} argue that some of the clumps formed in the disk may evolve to zero-age main sequence stars, but that most of these clumps can migrate inward and merge with the central protostar.

The temperature evolution of the gas cloud (shown in panel~B of Fig.~\ref{fig:prfsph}) is very similar in all simulations. In the final stage displayed in the plots, the outer layers of the cloud are at a temperature of \about \SI{8000}{K}. Further inwards, the temperature evolves nearly isothermally, though still gradually drops to about \SI{4000}{K}, until reaching a radius of about \SI{3e-6}{pc}. Inside this radius, the evolution proceeds nearly adiabatically and the temperature reaches \about \SI{7e4}{K} by the time the peak density is reached. This behaviour is expected based on the one-zone calculations, and for a more detailed description of the involved processes, see Sect.~\ref{subsec:one-zone}. 

The molecular hydrogen number fraction (shown in panel~C) exhibits some variation between simulations at larger radii, but converges for small radii, with only the T40R0 case deviating slightly from the others. At large radii, corresponding to low densities, the fraction slowly decreases as \ce{H2} is dissociated by the UV background. Next, a steep increase occurs at the radius corresponding to a density of \SI{e10}{cm^{-3}}, due to 3-body formation becoming efficient, after which formation and dissociation approximately balance each other for a broad density range. For the highest densities, where the gas is heating up, collisional dissociation starts to dominate and the fraction drops drastically. 
The \ce{H2} number fraction never gets much larger than \num{e-3}, in agreement with the one-zone test, which means that there is never enough \ce{H2} for molecular cooling to be important. 

The overall evolution of the electron number fraction (shown in panel~D) is again quite similar to what is expected from one-zone calculations. The T80R10 case deviates slightly from the others, in that the electron fraction starts to increase already at somewhat larger radii. This is again due to some mass buildup around that radius, reaching slightly higher temperatures than the surrounding matter.
The T20R10 case deviates quite strongly around the radius where for the other simulations the minimum occurs, which is due to the aforementioned second mass concentration at that radius.

The RMS turbulent velocity (shown in the panel~E) increases slowly with radius from \about \SI{1}{km/s} to \about \SI{10}{km/s} for all simulations. It is interesting that although initially the amount of turbulence is varied, later in the runs this difference is smoothed out and at least in the turbulent velocities there is no longer a clear difference between the high and low turbulence cases. 
The radial velocity (shown in panel~F) is similar for all simulations as well, and stays around \about \SI{11}{km/s} throughout most of the cloud. Only for the smallest radii, inside the minimum Jeans radius, does the radial velocity decrease down to \SIrange{1}{0.1}{km/s}, due to the pressure support in the clump.

The radial accretion rate (the rate of mass flow towards the central clump; shown in panel~H) is calculated as $\diff M/ \diff r = 4 \pi r^2 \rho v_{rad}$, where $\rho$ is the density and $v_{rad}$ the radial velocity. The rate varies somewhat between different simulations, although there does not seem to be a trend with either turbulence or rotation.  
The large peak in the accretion rate for the T20R10 run around \about \SI{4e-5}{pc} is due to the close connection of the second mass concentration to the central clump, locally boosting the accretion rate. It can be seen that both the density and radial velocity show a peak at the same location, causing the enhanced accretion rate. Similar features in the accretion rate were found by \citet{2009MNRAS.396..343R}, who also attribute them to clumps of high-density gas.
Overall accretion rates of a few solar masses per year are observed in all cases. Given such high accretion rates, a supermassive star of \SI{e5}{\Msun} is expected to form within \SI{e5}{years}.

From the density projections (Figs.~\ref{fig:prjx} and \ref{fig:prjz}), it can be seen that for the simulations including rotation a disk has formed. Stronger rotation leads to a flatter, more extended disk, and more pronounced spiral structures. 
The differences in turbulent structures show on the \SI{0.1}{pc} scale, with an increased amount of structure for higher initial turbulent velocities, also enhanced by stronger rotation. On smaller scales, the differences are no longer clear, as can also be seen from the turbulent velocity profiles in panel~E of Fig.~\ref{fig:prfsph}.

\begin{figure*}[!htbp]
 \centering
   \includegraphics[width=17cm]{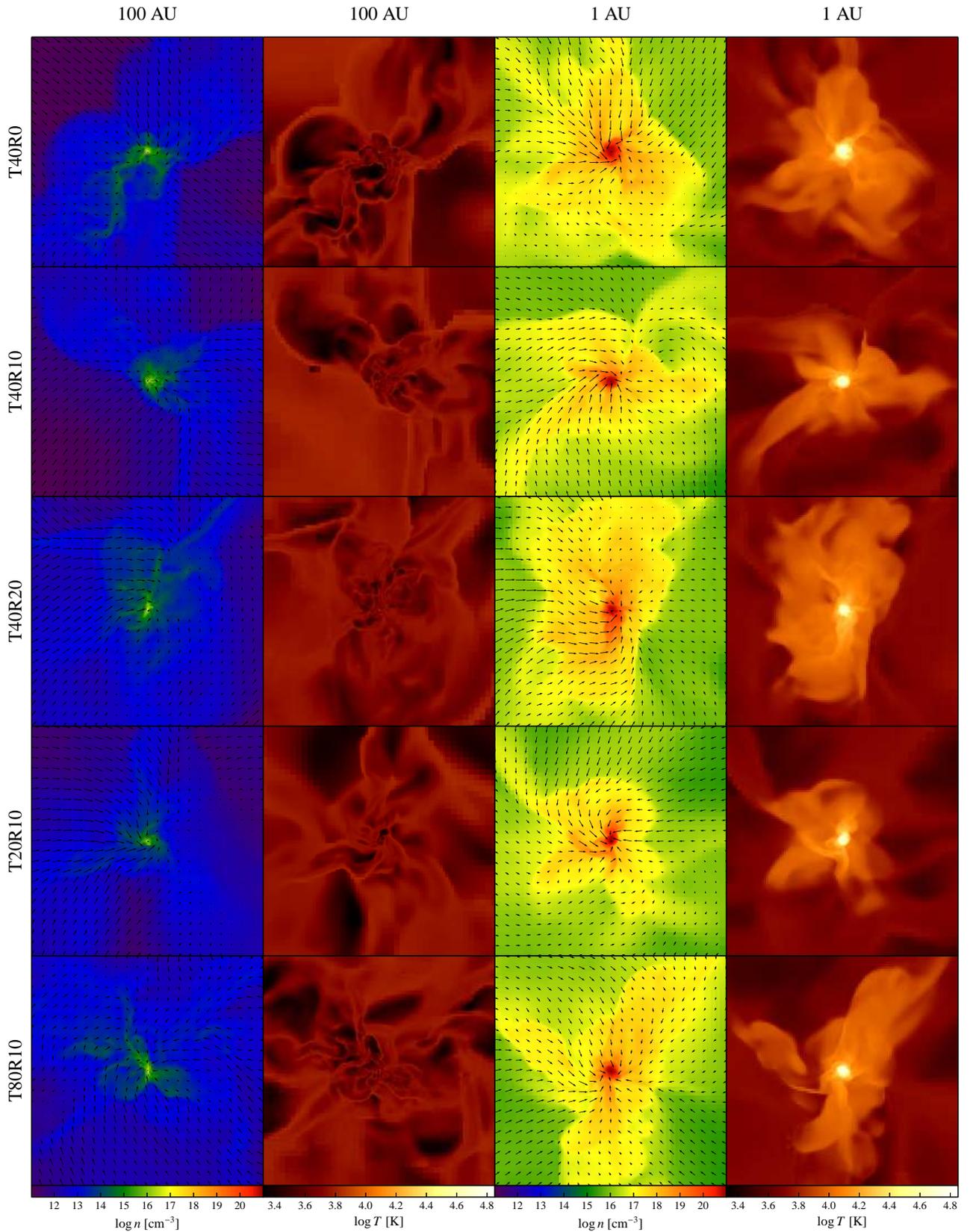}
 \caption{Density and temperature slices along the x-axis for all simulations and for two different scales at the peak density output. The simulation details and abbreviations are listed in Table~\ref{tab:simpars}.}
\label{fig:slcx}
\end{figure*}

\begin{figure*}[!htbp]
 \centering
   \includegraphics[width=17cm]{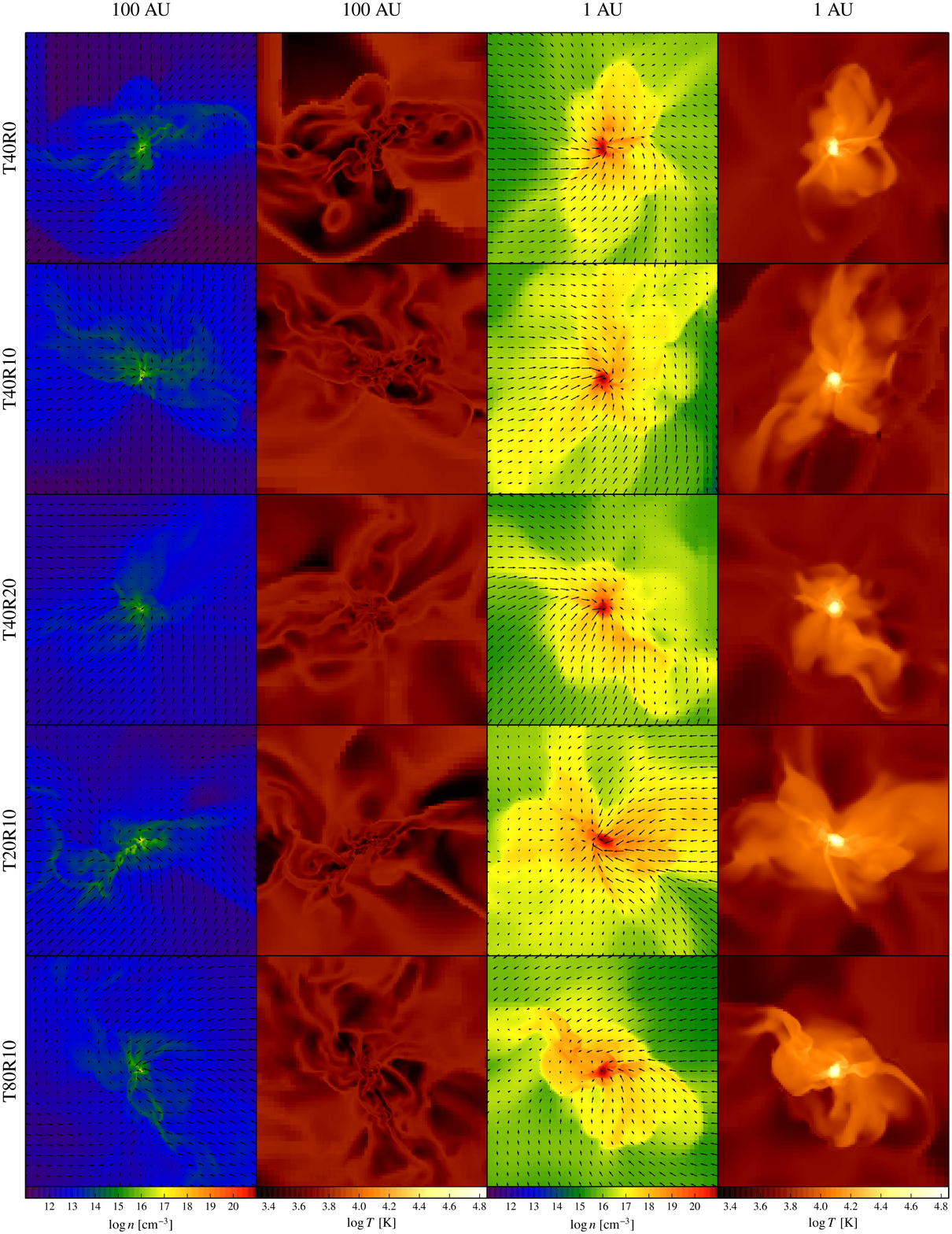}
 \caption{Density and temperature slices along the z-axis for all simulations and for two different scales at the peak density output. The simulation details and abbreviations are listed in Table~\ref{tab:simpars}.}
\label{fig:slcz}
\end{figure*}

Figs.~\ref{fig:slcx} and \ref{fig:slcz} show temperature slices for two different scales, next to density slices of the same area. It can be seen that there are hot regions of gas surrounding slightly cooler patches. 
Such warmer and cooler patches result from local compression and expansion of the gas due to turbulent motions.

\begin{figure*}[!htbp]
 \centering
   \includegraphics[width=17cm]{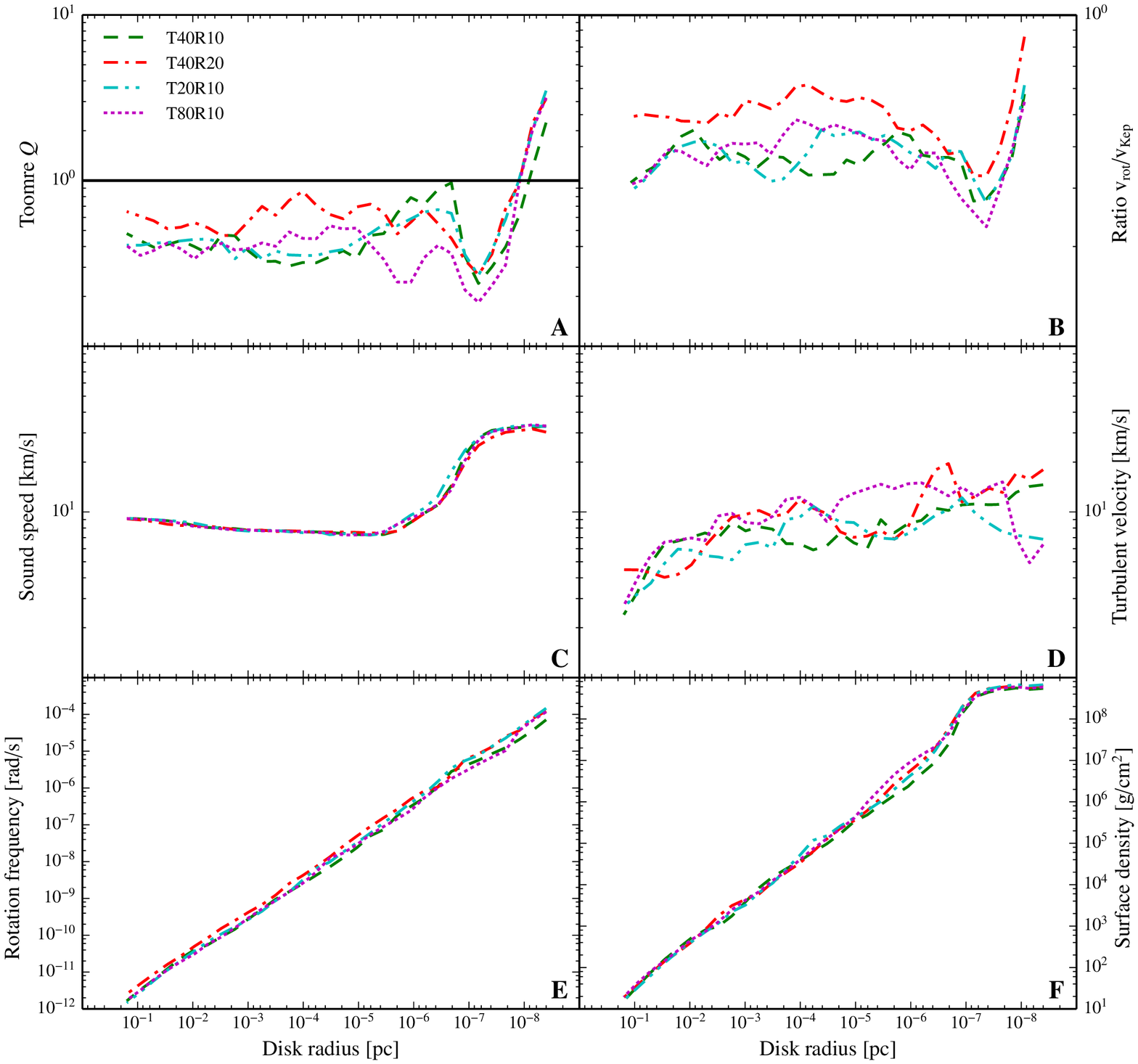}
 \caption{Physical quantities, weighted by mass, disk averaged and radially binned, as a function of radius at the peak density output for the different simulations. A: Toomre Q parameter; B: ratio of the rotational to the Keplerian velocity; C: sound speed; D: turbulent velocity; E: rotational frequency; F: surface density. The simulation details and abbreviations are listed in Table~\ref{tab:simpars}.}
\label{fig:prfdsk}
\end{figure*}

In Fig.~\ref{fig:prfdsk} we explore the properties of the disk, by displaying several disk averaged, radially binned quantities (using the radius in the x-y plane) for all simulations except T40R0 (as there is no disk present), centred on the peak density location (hereafter referred to as the central clump). The data shown has been obtained at the end of each simulation, when a peak density of \SI{e21}{cm^{-3}} was reached.

In panel~A, the Toomre $Q$ parameter is shown. This parameter is calculated as $Q = \frac{\sigma \Omega}{\pi G \Sigma}$, where $\sigma$ is the RMS of the sound speed and the turbulent velocity (as both thermal and turbulent motions will play a role in stabilizing the disk; plotted in panel~C and D), $\Omega$ is the rotation frequency (plotted in panel~E), $G$ is the gravitational constant, and $\Sigma$ is the surface density (plotted in panel~F). The disk is stable when $Q$ is larger than a critical value, which is of order unity; when $Q$ approaches this threshold, the disk will become gravitationally unstable.
It can be seen that the disk is mildly unstable for all simulations. Only for the smallest radii $Q$ becomes decidedly larger than one, which is due to the close-to-adiabatic core in the central region. On the smallest scales, the adiabatic heating thus stabilizes the protostar against further collapse. We note that the adiabatic equation of state is however only an approximation, while real systems may evolve further via Kelvin-Helmholtz contraction.

The ratio of the rotational velocity to the Keplerian velocity (shown in panel~B) follows roughly the same behaviour as the Toomre $Q$ parameter. The ratio is more or less constant over most radii. Some imprint of the initial amount of rotation remains, as the T40R20 run has the highest ratio over nearly all radii. However, for all runs the ratio has increased compared to the initial value, as a spin-up occurs during collapse.

It is interesting to note that we do not find a clear trend with either turbulence or rotation in any of the measured quantities on the smaller scales. It appears that whatever the initial conditions are, at later stages the initial difference in turbulence and rotation is smoothed out on these scales. Of course, on larger scales the presence and size of a disk does vary according to the initial amount of rotation, and on intermediate scales there are more turbulent structures for an increasing amount of initial turbulence, but this does not affect the overall evolution of density, temperature, accretion rate, and other quantities on scales smaller than \SI{1}{pc}. 

Whether one or more clumps are present does not depend on the initial amount of turbulence or rotation either, as we have concluded from comparison runs with the same initial conditions and a different random seed, in which usually one, sometimes two, and in a single case three of these clumps form. However, we never find more than three clumps, none of which have collapsed as far as the main clump, meaning that there is not much fragmentation, irrespective of turbulence or rotation. 
As mentioned, the simulations evolved for another \about \SI{1.67e-2}{years} after the highest refinement level was reached. Given that no fragmentation occurs in most of our simulations during this time, it can be considered as a lower limit on the fragmentation time scale.

\subsubsection{The central object}
A quantification of the properties of the central clump in each simulation is listed in Table~\ref{tab:clumps}. We find only one of such collapsed clumps in each simulation. The location of the `knee' in the enclosed mass profile is taken as the clump radius. The mass enclosed in this radius corresponds approximately to the minimum Jeans mass (see also panel~G in Fig.~\ref{fig:prfsph}), and thus the clumps are gravitationally bound.
This object marks the onset of protostar formation. Due to computational constraints simulations we cannot evolve the simulations further, though we expect the gas in the surroundings to collapse to form a massive protostar.

\begin{table}[!htbp]
\caption{Properties of the central bound clump found in each simulation.}
\label{tab:clumps}
\centering
\begin{tabular}{lcc} 
	\hline\hline
	\multicolumn{3}{c}{\textbf{Clumps}} \\[3pt]	
	Run & Radius [pc / AU] & Mass [\Msun] \\ 
	\hline
	T40R0  & \num{9e-8} / \num{2e-2} & \num{7.7e-2} \\
	T40R10 & \num{9e-8} / \num{2e-2} & \num{6.8e-2} \\
	T40R20 & \num{9e-8} / \num{2e-2} & \num{6.5e-2} \\
	T20R10 & \num{9e-8} / \num{2e-2} & \num{7.5e-2} \\
	T80R10 & \num{9e-8} / \num{2e-2} & \num{6.5e-2} \\
	\hline
\end{tabular}
\tablefoot{The location of the `knee' in the enclosed mass function is taken as the minimum clump radius; the mass enclosed in this radius corresponds approximately to the minimum Jeans mass.}
\end{table}

Given the radial accretion rates shown in Fig.~\ref{fig:prfsph}, a supermassive star of \SI{e5}{\Msun} is expected to form within \SI{e5}{years}, assuming that the gas reservoir to accrete from is large enough. If the accretion rate remains higher than \SI{e-2}{\Msun/yr}, \citet{2012ApJ...756...93H} found that the star will have a bloated envelope and lower surface temperatures, which inhibits the emission of ionizing radiation. In this case, radiative feedback will not be able to interfere with the accretion process.
If accretion rates higher than \SI{0.14}{\Msun/yr} can be maintained until the core has exhausted its hydrogen content through nuclear burning (after \about \SI{7e6}{yr}), it is likely that the core of the star will collapse into a black hole, resulting in a quasistar \citep{2013A&A...558A..59S}.

\section{Discussion \& Conclusions} \label{sec:conclusion}
We have performed 3D adaptive mesh refinement simulations using the ENZO code, simulating the formation of a protostar up to unprecedented high central densities of \SI{e21}{cm^{-3}}, and spatial scales of a few solar radii. To achieve this goal, we have employed the KROME package to improve the modelling of the chemical and thermal processes.
Particularly, we have investigated how the fragmentation behaviour of collapsing primordial gas in the presence of a strong Lyman-Werner radiation background is influenced by varying amounts of turbulence and rotation.

We found that in the runs including rotation, a mildly unstable disk forms on scales of \about \SI{0.5}{pc}, with a more extended disk for the stronger rotating case, run T40R20. On somewhat smaller scales, \about \SI{0.1}{pc}, the amount of turbulent structures increases with increasing initial turbulent velocities, as one would expect. However, on even smaller scales, $\lesssim \SI{0.01}{pc}$, the differences between the runs disappear, and radial profiles of the density, temperature, accretion rate, and other quantities are all very similar, with no dependence on the initial amount of turbulence or rotation.
The thermal evolution of all runs is consistent with the one-zone result from \citet{2001ApJ...546..635O}.

In each simulation we have found a single bound clump collapsed up to a density of \SI{e21}{cm^{-3}}, with a radius of \SI{2e-2}{AU} and a mass of \about \SI{7e-2}{\Msun}, corresponding to the minimum Jeans mass. This clump marks the onset of protostar formation. Given the observed accretion rates of a few solar masses per year, the protostar is expected to become a quasistar with a mass of \SI{e5}{\Msun} within \SI{e5}{years}, assuming a high accretion rate can be maintained. 
\citet{2014MNRAS.443.2410F} have derived a detailed prediction for the initial mass function (IMF) of the first massive black holes formed in atomic cooling haloes, combining the physics of SMS evolution and direct collapse black hole formation and growth with cosmological merger-tree simulations. They have found that in the case that minihaloes can form stars and pollute the gas, the IMF is bimodal and spans a broad mass range, $M \approx (\numrange{0.5}{20})\times\SI{e5}{\Msun}$; while in the case that they cannot form stars, the IMF spans a narrower range, $M \approx (\numrange{1}{2.8})\times\SI{e6}{\Msun}$. 
However, they predominantly consider larger scales (several kpc) on a longer-term evolution than the study presented in this paper, as their focus is on modelling the dynamics of halo mergers and the implications for accretion.

In a single run presented in this study (T20R10), the gas fragments into three clumps instead of one. 
From comparison runs with the same initial conditions and a different random seed for the realization of the turbulent velocity field, we have concluded that this fragmentation does not depend on the initial amount of rotation or turbulence, as usually one, sometimes two, and in a single case three clumps are found, though never more than three. 
Thus, we do not find much fragmentation, irrespective of turbulence or rotation. It is not clear whether the additional clumps will continue to collapse, or instead accrete onto the main clump, though based on a simplified model of fragmentation in the accretion disk around a protostar, \citet{2014arXiv1406.5058I} argue that most of these clumps can migrate inward and merge with the central protostar. The simulations have been evolved for another \about \SI{1.67e-2}{years} (\about 5 freefall times) after the highest refinement level was reached. As no fragmentation occurs in most of our simulations during this time, it can be considered as a lower limit on the fragmentation time scale.
To quantify the amount of fragmentation with greater certainty, the simulations should be evolved for longer, though our findings at least hint at the robustness of the direct collapse scenario.

For our simulations, we have used a Lyman-Werner background with a T5 spectrum. However, we expect to find the final result to be similarly independent of turbulence or rotation,
as long as the intensity of the UV background is above the critical value, regardless of the stellar spectrum.

Recently, \citet{2014arXiv1404.4630I} have done a similar simulation to attempt to resolve protostar formation, starting from equally simplified, though somewhat different initial conditions. They start from a marginally supported isothermal sphere with an initial density of \SI{e4}{cm^{-3}} and temperature of \SI{8000}{K}. The mass and radius of their cloud are slightly smaller than ours, \SI{1.17e5}{\Msun} and \SI{10.8}{pc}, respectively, though they are of the same order of magnitude. They also resolve the Jeans length by at least 64 grid cells, and their limiting resolution is \SI{0.1}{AU}.
There are a few differences between our chemical models. Concerning three-body rates, they do not take reaction \ce{H + H + H2} into account, and for reaction \ce{H + H + H} they use the rate by \citet{1987ApJ...318...32S}, while we have used the updated rate by \citet{2013ApJ...773L..25F}. 
Additionally, their opacity corrections at high density are calculated based on the Rosseland mean opacity, while we have used a different treatment for each cooling process, as described in Sect.~\ref{subsec:chemistry}. For atomic cooling, we consider opacity from Rayleigh scattering, and a fudge opacity to reduce cooling at high densities, in accordance with findings from previous one-zone studies. For \ce{H-} cooling, opacity from both Rayleigh scattering and \ce{H-} bound-free absorption is taken into account.

They have stopped their simulation when a temperature just in excess of \SI{e4}{K} was reached during the adiabatic phase, and find a hydrostatic core with a mass of \SI{1}{\Msun} and a radius of \SI{2}{AU}, at a peak density of \about \SI{5e16}{cm^{-3}}.
In our simulations, the onset of the adiabatic phase occurs approximately one order of magnitude in density later, likely due to differences in the chemical model, which results in our central clump being less massive, \about \SI{7e-2}{\Msun}, and smaller, \SI{2e-2}{AU}. 
Findings in agreement with ours include the resulting isothermal profile of the collapsing cloud, the fact that \ce{H2} cooling remains inefficient, and the accretion rate.
They did not include initial rotation, and do not find a disk, similar to our run without rotation (T40R0).

Presently, there are no simulations that resolve the formation of the protostar starting from cosmological initial conditions. 
However, our mass infall rates are in agreement with those found in cosmological direct collapse simulations reaching lower peak densities \citep{2013MNRAS.433.1607L,2013MNRAS.436.2989L}.

\citet{2008ApJ...682..745W} have performed cosmological simulations following the collapse of two atomic cooling haloes up to densities comparable to our peak density, though using a much simpler chemical model, neglecting, for example, \ce{H2} and \ce{H-} chemistry and cooling. Particularly, they did not consider optical depth effects, overestimating the cooling above column densities of \about \SI{e13}{cm^{-2}}, and thus did not obtain a transition towards an adiabatic equation of state. Therefore, the formation of a quasi-static object, like a protostar, was not observed. This can be clearly seen from their Fig.~5: there is no increase in temperature for the inner regions, nor a significant change in slope in the density or enclosed mass profiles. 
These important differences make it difficult to directly compare their results to ours. 
However, we can say that their radial velocity and density profiles are, except in the innermost region ($R < \SI{e-6}{pc}$), comparable to ours, meaning that the radial accretion rate should be of the same order of magnitude as well.

In this study, we did not take turbulence on subgrid scales into account \citep[for more details on subgrid scale turbulence, see e.g.][]{2006A&A...450..265S,2011A&A...528A.106S}. As has been shown by \citet{2013MNRAS.430..588L,2013MNRAS.433.1607L,2013MNRAS.436.2989L}, turbulence on unresolved scales affects the morphology of the collapsing gas.
Thus, it would be interesting to investigate whether this affects our results. 
Another caveat is the absence of magnetic fields. \citet{2013MNRAS.432..668L,2014MNRAS.440.1551L} have demonstrated that even a very small seed field can be effectively amplified by the small-scale dynamo mechanism. The resulting strong magnetic field provides additional support against gravity and helps to suppress fragmentation.
A further caveat is the simplification of the cooling functions, and more importantly, the opacities. Future work should include a more detailed treatment of optical depth effects, possibly even employing radiative transfer, although this will be computationally more expensive.
In the future, it would also be useful to implement equilibrium chemistry, as then it will be possible to follow the evolution for longer, and study the accretion onto the protostar in more detail. 
Additionally, simulations that resolve protostar formation starting from cosmological initial conditions are needed, to rule out possible effects caused by an idealized setup.

\begin{acknowledgements}
Computations described in this work were performed using the ENZO code (\url{http://enzo-project.org}), which is the product of a collaborative effort of scientists at many universities and national laboratories.
The simulation results are analysed using yt, a multi-code analysis toolkit for astrophysical simulation data \citep{2011ApJS..192....9T}.
CVB, DRGS and MAL acknowledge funding by the Deutsche Forschungsgemeinschaft (DFG) under grant SFB 963/1 (project A12). 
DRGS, SB, and CVB thank the DFG for funding and computing time via the Schwerpunktprogram SPP 1573 `Physics of the Interstellar Medium' (grant SCHL 1964/1 -- 1).
TG acknowledges the Centre for Star and Planet Formation funded by the Danish National Research Foundation.
\end{acknowledgements}

%% The Bibliography
\bibliographystyle{aa}
{\footnotesize \bibliography{allrefs}}

\onecolumn
\appendix
  \section{Reaction rates} \label{app:rates}
  %%Appendix table with reactions

% New counter for referencing
\newcounter{raterefno}
\newcommand{\ratereflabel}[1]{\refstepcounter{raterefno}\label{#1}}

\renewcommand{\arraystretch}{1.8}
\begin{longtab}
\small
%\centering
\begin{longtable}[t]{llll} %when equations fixed, p not necessary
	\caption{\label{tab:rates} Reaction rate coefficients.} \\
	\hline\hline %\toprule
	Reaction & Rate coefficient (\si{cm^{3}.s^{-1}}) & Range & Ref. \\ 
	\hline %\midrule
	\endfirsthead
	
	%\multicolumn{4}{c}{{\bfseries \tablename\ \thetable{}} -- Continued from previous page} \\
	%\midrule
	\caption{continued.}\\
	\hline\hline
	Reaction & Rate coefficient (\si{cm^{3}.s^{-1}}) & Range & Ref. \\ 
	\hline  %\midrule
	\endhead
	\hline
	\endfoot
	
	\ce{H + e- $\,\rightarrow\,$ H+ + 2e-} 
	& $\begin{aligned}[t] k_1 = &\exp\Big[-32.71396786 \\
		&+ 13.5365560 \ln{T_{eV}} \\
		&- 5.73932875 (\ln{T_{eV}})^2 \\
		&+ 1.56315498 (\ln{T_{eV}})^3 \\
		&- 0.28770560 (\ln{T_{eV}})^4 \\
		&+ \num{3.48255977e-2} (\ln{T_{eV}})^5 \\
		&- \num{2.63197617e-3} (\ln{T_{eV}})^6 \\
		&+ \num{1.11954395e-4} (\ln{T_{eV}})^7 \\
		&- \num{2.03914985e-6} (\ln{T_{eV}})^8\Big] \end{aligned}$  
	& & \ref{r1} \\
	
	\ce{H+ + e- $\,\rightarrow\,$ H + $\gamma$} 
	& $\begin{aligned}[t] k_2 = \num{3.925e-13} T_{eV}^{-0.6353} \end{aligned}$ 
	& $T \leq \SI{5500}{K}$ & \ref{r2} \\

	& $\begin{aligned}[t] k_2 = &\exp\Big[-28.61303380689232 \\
		&- 0.7241125657826851 \ln{T_{eV}} \\
		&- 0.02026044731984691 (\ln{T_{eV}})^2 \\
		&- 0.002380861877349834 (\ln{T_{eV}})^3 \\
		&- 0.0003212605213188796 (\ln{T_{eV}})^4 \\
		&- 0.00001421502914054107 (\ln{T_{eV}})^5 \\
		&+ \num{4.989108920299513e-6} (\ln{T_{eV}})^6 \\
		&+ \num{5.755614137575758e-7} (\ln{T_{eV}})^7 \\
		&- \num{1.856767039775261e-8} (\ln{T_{eV}})^8 \\
		&- \num{3.071135243196595e-9} (\ln{T_{eV}})^9\Big] \end{aligned}$ 
	& $T > \SI{5500}{K}$ & \ref{r2} \\
	
	\ce{He + e- $\,\rightarrow\,$ He+ + 2e-} 
	& $\begin{aligned}[t] k_3 = &\exp\Big[-44.09864886 \\
		&+ 23.91596563 \ln{T_{eV}} \\
		&- 10.7532302 (\ln{T_{eV}})^2 \\
		&+ 3.05803875 (\ln{T_{eV}})^3 \\
		&- 0.56851189 (\ln{T_{eV}})^4 \\
		&+ \num{6.79539123e-2} (\ln{T_{eV}})^5 \\
		&- \num{5.00905610e-3} (\ln{T_{eV}})^6 \\
		&+ \num{2.06723616e-4} (\ln{T_{eV}})^7 \\
		&- \num{3.64916141e-6} (\ln{T_{eV}})^8 \Big] \end{aligned}$
	& & \ref{r1} \\
	
	\ce{He+ + e- $\,\rightarrow\,$ He + $\gamma$} 
	& $\begin{aligned}[t] k_4 =\, &\num{3.925e-13} T_{eV}^{-0.6353} \\
		&+ \num{1.544e-9} T_{eV}^{-1.5} \Big[0.3 \exp{\left(-48.596/T_{eV}\right)} \\
		&+ \exp{\left(-40.496/T_{eV}\right)} \Big] \end{aligned}$ 
	& & \ref{rr4a}, \ref{rr4b} \\
	
	\ce{He+ + e $\,\rightarrow\,$ He++ + 2e-} 
	& $\begin{aligned}[t] k_5 = &\exp\Big[-68.71040990212001 \\
		&+ 43.93347632635 \ln{T_{eV}} \\
		&- 18.48066993568 (\ln{T_{eV}})^2 \\
		&+ 4.701626486759002 (\ln{T_{eV}})^3 \\
		&- 0.7692466334492 (\ln{T_{eV}})^4 \\
		&+ 0.08113042097303 (\ln{T_{eV}})^5 \\
		&- 0.005324020628287001 (\ln{T_{eV}})^6 \\
		&+ 0.0001975705312221 (\ln{T_{eV}})^7 \\
		&- \num{3.165581065665e-6} (\ln{T_{eV}})^8 \Big] \end{aligned}$
	& $T > \SI{9280}{K} $ & \ref{r2} \\
	
	\ce{He++ + e- $\,\rightarrow\,$ He+ + $\gamma$} 
	& $\begin{aligned}[t] k_6 = \, &\num{3.36e-10} T^{-1/2} (T/1000)^{-0.2} \\ 
		&(1 + (\num{e-6} T)^{0.7})^{-1} \end{aligned}$
	& & \ref{r3} \\
	
	\ce{H + e- $\,\rightarrow\,$ H- + $\gamma$} 
	& $\begin{aligned}[t] k_7 = \num{6.775e-15} T_{eV}^{0.8779} \end{aligned}$ 
	& & \ref{rr7} \\
	%it's a fit that agrees with other fits to the data in De Jong 1972, but this specific fit only appears
	%in ENZO and no reference is given there (other papers also took it from ENZO, like Shang et al. 2010)
	
	\ce{H- + H $\,\rightarrow\,$ H2 + e-} 
	& $\begin{aligned}[t] k_8 = \num{1.43e-9} \end{aligned}$ 
	& $T \leq \SI{1160}{K}$ & \ref{r2} \\
	
	& $\begin{aligned}[t] k_8 = &\exp\Big[-20.06913897587003 \\
		&+ 0.2289800603272916 \ln{T_{eV}} \\
		&+ 0.03599837721023835 (\ln{T_{eV}})^2 \\
		&- 0.004555120027032095 (\ln{T_{eV}})^3 \\
		&- 0.0003105115447124016 (\ln{T_{eV}})^4 \\
		&+ 0.0001073294010367247 (\ln{T_{eV}})^5 \\
		&- \num{8.36671960467864e-6} (\ln{T_{eV}})^6 \\
		&+ \num{2.238306228891639e-7} (\ln{T_{eV}})^7 \Big] \end{aligned}$
	& $T > \SI{1160}{K}$ & \ref{r2} \\
	
	\ce{H + H+ $\,\rightarrow\,$ H2+ + $\gamma$} 
	& $\begin{aligned}[t] k_9 = \num{1.85e-23} T^{1.8} \end{aligned}$ 
	& $T \leq \SI{6700}{K}$ & \ref{r4} \\
	
	& $\begin{aligned}[t] k_9 = \num{5.81e-16} (T/\num{56200})^{(-0.6657*\log{(T/\num{56200})})} \end{aligned}$ 
	& $T > \SI{6700}{K}$ & \ref{r4} \\
	
	\ce{H2+ + H $\,\rightarrow\,$ H2 + H+} 
	& $\begin{aligned}[t] k_{10} = \num{6.0e-10} \end{aligned}$ 
	& & \ref{r5} \\
	
	\ce{H2 + H+ $\,\rightarrow\,$ H2+ + H} 
	& $\begin{aligned}[t] k_{11} = &\exp\Big[-24.24914687731536 \\
		&+ 3.400824447095291 \ln{T_{eV}} \\
		&- 3.898003964650152 (\ln{T_{eV}})^2 \\
		&+ 2.045587822403071 (\ln{T_{eV}})^3 \\
		&- 0.5416182856220388 (\ln{T_{eV}})^4 \\
		&+ 0.0841077503763412 (\ln{T_{eV}})^5 \\
		&- 0.007879026154483455 (\ln{T_{eV}})^6 \\
		&+ 0.0004138398421504563 (\ln{T_{eV}})^7 \\
		&- \num{9.36345888928611e-6} (\ln{T_{eV}})^8 \Big] \end{aligned}$
	& & \ref{r2} \\
	
	\ce{H2 + e- $\,\rightarrow\,$ 2H + e-} 
	& $\begin{aligned}[t] k_{12} = \num{5.6e-11} \exp{(-102124/T)} T^{1/2} \end{aligned}$ 
	& & \ref{r6} \\
	
	\ce{H2 + e- $\,\rightarrow\,$ H + H-} 
	& $\begin{aligned}[t] k_{13} = 36.7 T^{-2.28} \exp{(-47172/T)} \end{aligned}$ 
	& & \ref{r7} \\
	
	\ce{H2 + H $\,\rightarrow\,$ 3H} 
	& See expression in [\ref{r8}]
	& & \ref{r8} \\
	
	\ce{H2 + H2 $\,\rightarrow\,$ H2 + 2H} 
	& $\begin{aligned}[t] &k_{15} = \dex\Big[\left(n_H/n_{cr} \left(1 + n_H/n_{cr}\right)^{-1}\right) \log{k_{15,LTE}} \\ 
		&\qquad \> + \left(1 + n_H/n_{cr}\right)^{-1} \log{k_{15,v0}}\Big] \\
		&k_{15,v0} = \left(\num{6.0465e-30} T^{4.1881}\right) \\ 
		&\qquad \> / \left(1 + \num{6.7606e-6} T\right)^{5.6881} \\
		&\qquad \> \exp{\left(-54657.4/T\right)} \\
		&k_{15,LTE} = \num{1.3e-9} \exp{\left(-53300/T\right)} \\
		&\text{See Section~\ref{subsubsec:chemhc} for } n_{cr} \end{aligned}$ 
	& & \ref{r9}, \ref{r4} \\
	
	\ce{H- + e- $\,\rightarrow\,$ H + 2e-} 
	& $\begin{aligned}[t] k_{16} = &\exp\Big[-18.01849334273 \\
		&+ 2.360852208681 \ln{T_{eV}} \\
		&- 0.2827443061704 (\ln{T_{eV}})^2 \\
		&+ 0.01623316639567 (\ln{T_{eV}})^3 \\
		&- 0.03365012031362999 (\ln{T_{eV}})^4 \\
		&+ 0.01178329782711 (\ln{T_{eV}})^5 \\
		&- 0.001656194699504 (\ln{T_{eV}})^6 \\
		&+ 0.0001068275202678 (\ln{T_{eV}})^7 \\
		&- \num{2.631285809207e-6} (\ln{T_{eV}})^8 \Big]  \end{aligned}$
	& & \ref{r1} \\
	
	\ce{H- + H $\,\rightarrow\,$ 2H + e-} 
	& $\begin{aligned}[t] k_{17} = \num{2.5634e-9} T_{eV}^{\num{1.78186}} \end{aligned}$ 
	& $T \leq \SI{1160}{K} $ & \ref{r2} \\
	
	& $\begin{aligned}[t] k_{17} = &\exp\Big[-20.37260896533324 \\
		&+ 1.139449335841631 \ln{T_{eV}} \\
		&- 0.1421013521554148 (\ln{T_{eV}})^2 \\
		&+ 0.00846445538663 (\ln{T_{eV}})^3 \\
		&- 0.0014327641212992 (\ln{T_{eV}})^4 \\
		&+ 0.0002012250284791 (\ln{T_{eV}})^5 \\
		&+ 0.0000866396324309 (\ln{T_{eV}})^6 \\
		&- 0.00002585009680264 (\ln{T_{eV}})^7 \\
		&+ \num{2.4555011970392e-6} (\ln{T_{eV}})^8 \\
		&- \num{8.06838246118e-8} (\ln{T_{eV}})^9 \Big] \end{aligned}$ 
	& $T > \SI{1160}{K} $ & \ref{r2} \\
	
	\ce{H- + H+ $\,\rightarrow\,$ 2H} 
	& $\begin{aligned}[t] k_{18} = \num{6.5e-9} T_{eV}^{-1/2} \end{aligned}$ 
        & & \ref{r10} \\

	\ce{H- + H+ $\,\rightarrow\,$ H2+ + e-} 
	& $\begin{aligned}[t] k_{19} = \num{4e-4} T^{-1.4} \exp{\left(-15100/T\right)} \end{aligned}$ 
	& $T \leq \SI{e4}{K} $ & \ref{rr19a} \\

	& $\begin{aligned}[t] k_{19} = \num{e-8} T^{-0.4} \end{aligned}$ 
	& $T > \SI{e4}{K} $ & \ref{rr19b} \\

	\ce{H2+ + e- $\,\rightarrow\,$ 2H} 
	& $\begin{aligned}[t] k_{20} = \num{e-8} \end{aligned}$ 
	& $T \leq \SI{617}{K}$ & \ref{r2} \\

	& $\begin{aligned}[t] k_{20} = \num{1.32e-6} T^{-0.76} \end{aligned}$ 
	& $T > \SI{617}{K}$ & \ref{r2} \\

	\ce{H2+ + H- $\,\rightarrow\,$ H + H2} 
	& $\begin{aligned}[t] k_{21} = \num{5.0e-6} T^{-1/2} \end{aligned}$ 
	& & \ref{r10} \\

	\ce{H + H + H $\,\rightarrow\,$ H2 + H} 
	& $\begin{aligned}[t] k_{22} = \num{6e-32} T^{-1/4} + \num{2e-31} T^{-1/2} \end{aligned}$ 
	& & \ref{r11} \\ %Forrey 2013

	\ce{H + H + H2 $\,\rightarrow\,$ 2H2} 
	& $\begin{aligned}[t] k_{23} = \left(\num{6e-32} T^{-1/4} + \num{2e-31} T^{-1/2}\right)/8 \end{aligned}$ 
	& & \ref{r11}, \ref{r12} \\

	\ce{H- + $\gamma$ $\,\rightarrow\,$ H + e-} 
	& $\begin{aligned}[t] k_{24} &= \num{e-10} \alpha J_{21} \\ 
		\alpha &= 2000 \text{ for T4 spectrum} \\ 
		\alpha &= 0.1 \text{ for T5 spectrum} \end{aligned}$ 
	& & \ref{rr19b} \\
	
	\ce{H2 + $\gamma$ $\,\rightarrow\,$ 2H} 
	& $\begin{aligned}[t] k_{25} &= \num{e-12} \beta J_{21} \\ 
		\beta &= 3 \text{ for T4 spectrum} \\
		\beta &= 0.9 \text{ for T5 spectrum} \end{aligned}$ 
	& & \ref{rr19b} \\

	%\bottomrule
\end{longtable}
\tablefoot{$T$ and $T_{eV}$ are the gas temperature in units of K and eV, respectively.}
\tablebib{ 
\protect\ratereflabel{r1}[\ref{r1}] \citet{1987ephh.book.....J};
\protect\ratereflabel{r2}[\ref{r2}] \citet{1997NewA....2..181A};
\protect\ratereflabel{rr4a}[\ref{rr4a}] \citet{1992ApJS...78..341C};
\protect\ratereflabel{rr4b}[\ref{rr4b}] \citet{1973A&A....25..137A};
\protect\ratereflabel{r3}[\ref{r3}] \citet{2000ApJ...534..809O};
\protect\ratereflabel{rr7}[\ref{rr7}] \citet{1972A&A....20..263D};
\protect\ratereflabel{r4}[\ref{r4}] \citet{1987ApJ...318...32S};
\protect\ratereflabel{r5}[\ref{r5}] \citet{1979JChPh..70.2877K};
\protect\ratereflabel{r6}[\ref{r6}] \citet{1991ApJ...383..511D};
\protect\ratereflabel{r7}[\ref{r7}] \citet{2007A&A...470..811C};
\protect\ratereflabel{r8}[\ref{r8}] \citet{1996ApJ...461..265M};
\protect\ratereflabel{r9}[\ref{r9}] \citet{1998ApJ...499..793M};
\protect\ratereflabel{rr19a}[\ref{rr19a}] \citet{1978JPhB...11L.671P};
\protect\ratereflabel{rr19b}[\ref{rr19b}] \citet{2010MNRAS.402.1249S};
\protect\ratereflabel{r10}[\ref{r10}] \citet{1987IAUS..120..109D};
\protect\ratereflabel{r11}[\ref{r11}] \citet{2013ApJ...773L..25F};
\protect\ratereflabel{r12}[\ref{r12}] \citet{1983ApJ...271..632P}.
}
\end{longtab}

\end{document}